\documentclass[lettersize,journal]{IEEEtran}
\pdfoutput=1
\usepackage{graphicx}
\usepackage{subfiles}
\usepackage[caption=false,font=footnotesize,labelfont=sf,textfont=sf]{subfig}
\usepackage{amsmath}
\usepackage{amssymb}
\usepackage[thmmarks,amsmath]{ntheorem} 
\usepackage{blkarray}
\usepackage{mathtools}
\usepackage{xcolor}
\usepackage{colortbl}
\usepackage{tikz}
\usepackage{tabularx}
\usepackage[caption=false]{subfig}
\usepackage[T1]{fontenc}
\usepackage[utf8]{inputenc}
\usepackage{multirow}
\usepackage{booktabs}
\usepackage[ruled,vlined]{algorithm2e}

\SetAlgoNlRelativeSize{-1} 
\usepackage{algpseudocode}
\newtheorem{definition}{Definition}

\usepackage{nomencl}
\makenomenclature

\usepackage{etoolbox}
\renewcommand\nomgroup[1]{%
  \item[\bfseries
  \ifstrequal{#1}{A}{Acronyms}{%
  \ifstrequal{#1}{S}{Symbols}}%
]}

\begin{document}

\title{\vspace{-0.6cm}
Data-Driven Decision Making for Enhancing Small-Signal Stability in Hybrid AC/DC Grids Through Converter Control Role Assignment}

\author{Francesca Rossi, Sergi Costa Dilmé, Josep Arévalo-Soler, Eduardo Prieto-Araujo,~\IEEEmembership{Senior Member,~IEEE},
and Oriol Gomis-Bellmunt,~\IEEEmembership{Fellow,~IEEE}
\vspace{-1.1cm}}



\maketitle

\begin{abstract}
Hybrid AC/DC transmission grids incorporate Modular Multilevel Converters functioning as Interconnecting Power Converters (IPCs). The control role assigned to each converter significantly influences grid dynamics. Traditionally, these converters operate with static control roles, but recent studies have proposed scheduling their roles based on day-ahead forecasts to enhance stability performance. However, in systems with high renewable energy penetration, forecast deviations can render scheduled control assignments suboptimal or even lead to instability. To address this challenge, this work proposes an online scheduling recalculation algorithm that dynamically adapts IPC control roles during system operation. The approach leverages a data-driven multi-criteria decision-making framework, integrating surrogate models of conventional small-signal stability analysis tools to enable a fast computation of system stability and stability performance indicators.
\end{abstract}

\begin{IEEEkeywords}
Small-signal stability, Machine Learning, HVDC, Control Role Assignment
\vspace{-0.6cm}
\end{IEEEkeywords}
\section{Introduction}
\vspace{-0.1cm}
Transmission 
systems are increasingly incorporating High-Voltage Direct-Current (HVDC) technologies, which play a crucial role in integrating offshore wind power plants and enabling the efficient transmission of electricity over long distances~\cite{ramosready4dc}. Numerous projects 
are currently being developed, encompassing both point-to-point and multi-terminal connections 
~\cite{ramosready4dc}. 
Nowadays, 
Modular Multilevel Converter (MMC) technology has become the dominant solution for the Interconnecting Power Converters (IPCs). A key feature of MMCs is the ability to control voltage at both the AC and DC terminals. Therefore, in addition to the conventional grid-following (GFL) control, 
MMCs can also operate in grid-forming (GFM) mode on either the AC or DC side~\cite{9729620}. 

GFL control has been observed to experience instability in weak grid conditions, whereas GFM control can improve stability and provide support for grid operation in such scenarios~\cite{9714816}. However, high penetration of GFM converters may introduce oscillations in strong grids or lead to undesirable interactions when multiple GFM converters are located too close electrically~\cite{9714816}. The emergence of these converter control-driven issues affecting system dynamics has increased interest in studying how to assign appropriate control roles to ensure and enhance grid stability. In~\cite{10508461}, the required level of GFM penetration to ensure small-signal stability and strengthen a weak grid with multiple wind farms is analyzed. The study assumes that wind turbines are connected via GFL converters, while a storage system interfaced through a GFM converter compensates for wind power fluctuations and enhances grid strength. However, a static control assignment approach is adopted, without exploring different operating conditions. 
In~\cite{cui2024control}, the use of inverter-based resources (IBRs) for stability services is explored. The need for dedicated converters is bypassed by dynamically assigning control roles to the existing IBRs during operation. The optimal operating point (OP) and GFM allocation are obtained as solution of a Mixed Integer Linear Programming (MILP) Unit Commitment problem, considering frequency and small-signal stability constraints.\\
Focusing on hybrid AC/DC grids, the authors in~\cite{10017135,10443561} investigated the feasibility of applying control role scheduling to IPCs. As defined in~\cite{10443561}, the assignment of specific control roles to each IPC forms a Converter Control Role Configuration (CCRC). For a given system OP, 
the system dynamics varies depending on the applied CCRC, ranging from instability to good dynamic performance. Therefore, the need for dynamic CCRC scheduling arises primarily from the fact that no single CCRC guarantees stability across all operating conditions~\cite{10443561}. Additionally, among stable CCRCs, it is possible to select the one that provides the best dynamic performance. In~\cite{10443561}, numerical indicators are formulated to evaluate both steady-state and dynamic small-signal stability performance. Based on these indicators, an MILP optimization problem is developed to dynamically schedule IPC control role assignments, ensuring that, for each system operating point, the selected CCRC maximizes stability performance.

The scheduling approaches presented in~\cite{cui2024control,10443561} have been designed for day-ahead operation and thus rely on accurate day-ahead forecasts of demand and generation. However, significant discrepancies between actual renewable generation or demand and the corresponding forecasts can make a predefined CCRC schedule suboptimal or even unstable for the actual OP. This highlights the need for a fast recalculation tool, capable of determining the most suitable CCRC assignment during online operation. To accelerate computation, this work proposes two key strategies: (i) replacing the MILP-based approach with a Multi-Criteria Decision-Making (MCDM) algorithm and (ii) substituting the mathematically exact small-signal stability models used for stability assessment and performance indicator calculation with their data-driven surrogate models. MILP is not considered a viable approach due to its significant computational requirements, which render it impractical for real-size problems~\cite{floudas2008encyclopedia}. In contrast, MCDM exhibits polynomial computational complexity, making it a more scalable and suitable alternative for solving large-size problems~\cite{floudas2008encyclopedia}. The use of surrogate models to reduce computational time is explored~\cite{BHOSEKAR2018250}. The stability assessment is performed by a classification algorithm and regressions are trained to estimate the stability indicators.

The main contributions of this paper are:  
\begin{itemize}  
    \item A data-driven MCDM algorithm for online CCRC assignment to IPCs in hybrid AC/DC grids, designed to enhance small-signal stability performance.  
    \item A comprehensive methodology for training data-driven surrogate models for small-signal stability assessment and the estimation of stability performance indicators.  
    \item A methodology to enhance the scalability of the MCDM approach by reducing the number of CCRCs alternatives, retaining only those necessary for optimal stability performance. This selection process leverages data mining techniques to cluster CCRCs with similar stability behavior within the same operating region and applies set intersection methods to integrate information from different stability indicators to support the selection.  
\end{itemize}

\vspace{-0.5cm}
\section{Small-signal Stability and Small-signal Stability Performance Indicators}
\label{sec: ss_ind}
\subsection{Conventional Exact Models} 
Small-signal stability assessment is conventionally based on the eigenvalues analysis of the linearized state-space of the system, evaluated around an equilibrium point \cite{kundur1994power}. Such equilibrium point can be obtained by power flow (PF) calculation or time-domain simulation. Following Lyapunov's first method for small-signal stability assessment, the system results asymptotically stable if all the eigenvalues ($\lambda$) have negative real parts \cite{kundur1994power}. A binary label $\Upsilon \in \{0,1\}$ can be used to indicate the outcome of the small-signal stability analysis:
\vspace{-0.2cm}
\begin{align}
      \left\{\begin{aligned}       
    \Upsilon&=1 \quad if \max\{\Re(\lambda_1),...,\Re(\lambda_N)\}<0\\
    \Upsilon&=0 \quad otherwise 
    \end{aligned}\right.
    \label{eq: stab_lab}
\end{align}
\vspace{-0.5cm}
\begin{definition}{\bf{(Exact small-signal stability assessment)}}\label{def: exact_ss}
The \textit{exact} computation of the small-signal stability is indicated as 
\vspace{-0.5cm}
\begin{equation}
    \Psi: (x,u,y) \longrightarrow \Upsilon
\vspace{-0.2cm}
\end{equation}
\noindent where $x$, $u$, $y$ are the states, the inputs, and the outputs of the system state-space, respectively. It is computed for a certain OP and CCRC.
\end{definition}
\vspace{-0.2cm}
When the system is stable, its small-signal stability performance can be evaluated by analyzing the behavior of the system's transfer function matrix, \( G(s) \). This matrix, of dimension \( \mathbb{R}^{|y| \times |u|} \), is defined as $G(s) = \frac{y(s)}{u(s)}$, where \( y(s) \) and \( u(s) \) represent the system's outputs and inputs in the Laplace domain, respectively, and \( |y| \) and \( |u| \) denote the number of outputs and inputs. To capture the specific dynamics that it is intended to enhance, $G(s)$ has to be formulated by properly setting the inputs and outputs signals. Then, to evaluate the dynamic performances, one possibility 
is to compute the $\mathcal{H}_2$-norm of the transfer functions matrix, $||G(s)||_2$. 
The $\mathcal{H}_2$-norm provides a unique scalar value that is 
an estimation of the total output signals energy~\cite{skogestad2005multivariable}. The higher the norm value, the larger the sum of the signals energy content and the larger their oscillations before converging to the steady state. 
\vspace{-0.2cm}
\begin{definition}{\bf{(Exact $\mathcal{H}_2$-norm calculation)}}\label{def: exact_h2}
The \textit{exact} computation of the $\mathcal{H}_2$-norm of the system output signals $y$ is indicated as 
\vspace{-0.2cm}
\begin{equation} \label{eq: exact_H2}
\Gamma_{\mathcal{H}_{2,y}}: (x,u,y) \longrightarrow \mathcal{H}_{2,y}
\vspace{-0.2cm}
\end{equation}
\end{definition}

Another approach for formulating a small-signal stability performance indicator is to compute the \textit{DC Gain} (\(\mathcal{K}\)) of the transfer function matrix \(G(s)\). For each individual transfer function \( g(s) \in G(s) \), the DC Gain can be obtained from its canonical formulation:
    $g(s) = \frac{k \left(1+\sum_{i=1}^{m} b_i s^i \right)}{s^r \left(1+\sum_{j=1}^{l} a_j s^j \right)}$.
In this formulation, the coefficient \( k \) represents the \textit{DC Gain} of \( g(s) \). It can be computed as $k = \lim_{t \to \infty} y(t) = \lim_{s \to 0} s^r g(s).$
Therefore, this indicator measures the deviation of the output signal from its reference value once the system has reached a steady state. A higher \textit{DC Gain} value corresponds to a larger signal deviation, indicating poorer stability performance.
The \textit{DC Gain} of \( G(s) \) is then defined as the maximum \( k \) value among all transfer functions in \( G(s) \), representing the maximum deviation experienced by any output signal \( y \).
\vspace{-0.2cm}
\begin{definition}{\bf{(Exact $\mathcal{K}$ calculation)}}\label{def: exact_dcgain}
The \textit{exact} computation of the DC gain $\mathcal{K}$ of the system output signals $y$ is indicated as 
\vspace{-0.2cm}
\begin{equation} \label{eq: exact_DCgain}
    \Gamma_{\mathcal{K}_{y}}: (x,u,y) \longrightarrow \mathcal{K}_{y}
\end{equation}
\end{definition}
\vspace{-0.2cm}

In summary, if the small-signal stability assessment states that the system response, for a certain OP and CCRC, is stable, the $\mathcal{H}_2$-norms and the \textit{DC Gain} of relevant signals can be calculated. Such quantities indicate stability performances. 
\vspace{-0.7cm}
\subsection{Data-driven Surrogate Models}
\vspace{-0.1cm}
The 
tools 
of Definitions \ref{def: exact_ss}-\ref{def: exact_dcgain} are referred to as \textit{exact} calculations as they are based on the mathematical 
formulation of the linear system state-space and system transfer functions. 
In this paper, it is proposed to train machine learning (ML) algorithms to serve as data-driven \textit{surrogate} for these \textit{exact} models. These \textit{surrogate} models aim to replicate the output of the \textit{exact} models, specifically the stability response and values of stability indicators, by utilizing classification and regression algorithms trained on suitable data. These data are generated by computing the \textit{exact} models across various OPs and CCRCs, as detailed in Section \ref{sec: datagen}. The computed instances and their corresponding outcomes are then organized into data sets, each tailored to a specific model for training. All data sets have the same structure: a single data set can be generically described as $\mathcal{D}=(X|z)$, where $X$ and $z$ are 
the input and target variables of the model to be trained. 

Concerning the data-driven \textit{surrogate} for the small-signal stability assessment, a unique model is trained over a single data set, collecting quantities and outcomes from all the computed instances, encompassing the application of all the CCRCs.   
The output variable is $\Upsilon$. As it is a binary variable, the \textit{surrogate} model to be trained has to perform a classification.
The input quantities, $X$, relate to the OP and CCRC of the computed instances, as in the \textit{exact} models' calculations, though they do not match the \textit{exact} models' inputs precisely. In the \textit{exact} model calculation, comprehensive knowledge of the system is required, including a detailed formulation of the control schemes implemented and the precise tuning values of the control parameters. On the contrary, \textit{surrogate} models do not require information on control schemes or parameter tunning, only the electrical characteristics of the system. A detailed description of the input quantities is provided next.
\vspace{-0.1cm}
\begin{itemize}
    \item The PF input quantities, $X_{PF}$. These quantities can be obtained through PF calculations based on the system's OP. Consider a generic power system and define $\mathcal{N}$ as the set of buses, $\mathcal{G}$ as the set of generators, $\mathcal{L}$ as the set of loads, and $\mathcal{C}$ as the set of IPCs. Moreover, consider $\mathcal{T}$ as the set collecting Thevenin's equivalents that might be used for modeling portions of the AC grids for simulation and computation purposes. Then, the PF quantities are 
        $X_{PF}=[(V_i,\theta_i)_{\forall i\in\mathcal{N}},(P_i,Q_i)_{\forall i\in\mathcal{G}},(P_i,Q_i)_{\forall i\in\mathcal{C}},
        (P_i,Q_i)_{\forall i\in\mathcal{L}},\\
        (P_i,Q_i)_{\forall i\in\mathcal{T}}]$,
    where $V$ and $\theta$ are the voltage module and phase angle, $P$ and $Q$ are the active and reactive power.
    \item The IPC internal measures input quantities, $X_{IPC}$. As identified in \cite{PRIETOARAUJO2017}, they are 
        $X_{IPC}=[(V_i^{AC},\theta_{V,i}^{AC},V_i^{diff},\theta_{V,i}^{diff},V_i^{sum},\theta_{V,i}^{sum},        I_i^{diff},\theta_{I,i}^{diff},\\I_i^{sum},\theta_{I,i}^{sum})_{\forall i\in\mathcal{C}}]$,  
    which includes, for the $i$-th converter, the module of the three-phase voltage on the AC side ($V_i^{AC}$), the modules of the applied three-phase differential ($V_i^{diff}$) and additive ($V_i^{sum}$) voltages, the modules of the differential ($I_i^{diff}$) and additive ($I_i^{sum}$) currents. The quantities $\theta$s are the corresponding phase angles. These quantities can be calculated once the PF is solved and known the electrical 
    modeling of the IPCs.
    
    \item The input quantities related to the CCRC assignment, $X_{C}$. Since variations in control parameter tuning for converters are not considered in this study — only changes in control roles are — CCRC is accounted for by encoding the control roles as categorical variables. The CCRC assignment is then expressed by a sequence of discrete variables, each identifying a converter control role (CCR), as follows 
    $    X_{C}=[(CCR_i)_{\forall i\in\mathcal{C}}]$.
\end{itemize}

Summarizing the above mentioned consideration, the data-driven \textit{surrogate} model for a small-signal stability assessment is defined as follows.
\vspace{-0.1cm}
\begin{definition}\textbf{(Data-driven surrogate of the small-signal stability assessment)}\label{def: dd_stab} The data-driven \textit{surrogate} model performing the small-signal stability assessment is a classification algorithm trained over a data set $\mathcal{D}_\Upsilon=(X_{PF},X_{IPC},X_{C}|\Upsilon)$. It maps the small-signal stability response of the system for a range of OPs and CCRCs, as expressed by 
\vspace{-0.2cm}
\begin{equation}
    \hat{\Psi}: (X_{PF},X_{IPC},X_{C}) \longrightarrow \hat{\Upsilon}
\vspace{-0.2cm}
\end{equation} 
Set a CCRC and for a certain OP, not necessarily belonging to the training data set $\mathcal{D}_\Upsilon$, $\hat{\Psi}$ calculates an estimation of the system stability, $\hat{\Upsilon}$. 
\end{definition}
\vspace{-0.1cm}
Concerning the \textit{surrogate} models for the calculation of the stability performances indicators, a different approach is used. These indicators vary widely in value across different CCRCs, making a single model for all CCRCs less accurate. Therefore, for each CCRC, a separate data set is built and a separate model is trained. Hence, the variables describing the CCRCs, $X_C$, do not need to be included in the training data sets and as input of these models. In summary, for the data-driven \textit{surrogate} models of the stability performances indicators, the following definitions are provided.
\vspace{-0.2cm}
\begin{definition}\textbf{(Data-driven \textit{surrogate} of the $\mathcal{H}_2$-norm calculation)}\label{def: dd_h2} The data-driven \textit{surrogate} model performing the $\mathcal{H}_2$-norm calculation of system outputs $y$, for the $i$-th CCRC, is a regression algorithm trained over a data set $\mathcal{D}_{\mathcal{H}_{2,y}}^{i}=(X_{PF},X_{IPC}|\mathcal{H}_{2,y})$.
It fits, for the $i$-th CCRC, the value of the indicator for a range of OPs. 
\vspace{-0.2cm}
\begin{equation}
\hat{\Gamma}_{\mathcal{H}_{2,y}}^{i}: (X_{PF},X_{IPC}) \longrightarrow \hat{\mathcal{H}}_{2,y}
\vspace{-0.2cm}
\end{equation}   
Given an OP not necessarily belonging to 
$\mathcal{D}_{\mathcal{H}_{2,y}}^{i}$, $\hat{\Gamma}_{\mathcal{H}_{2,y}}^{i}$ calculates an estimation of the stability performance indicator $\hat{\mathcal{H}}_{2,y}$, considering the application of the $i$-th CCRC.
\end{definition}
\vspace{-0.3cm}
\begin{definition}\textbf{(Data-driven \textit{surrogate} of the $\mathcal{K}$ calculation)} 
\label{def: dd_dcgain}
The data-driven \textit{surrogate} model performing the $\mathcal{K}$ calculation of the output signals $y$, for the $i$-th CCRC, is a regression algorithm trained over a data set $\mathcal{D}_{\mathcal{K}_y}^{i}=(X_{PF},X_{IPC}|\mathcal{K}_{y})$.
It fits, for the $i$-th CCRC, the value of the indicator for a range of OPs. 
\vspace{-0.3cm}
\begin{equation}
\hat{\Gamma}_{\mathcal{K}_{y}}^{i}: (X_{PF},X_{IPC}) \longrightarrow \hat{\mathcal{K}}_{y}
\vspace{-0.2cm}
\end{equation}   
Given an OP not necessarily belonging to 
$\mathcal{D}_{\mathcal{K}_y}^{i}$, $\hat{\Gamma}_{\mathcal{K}_{y}}$ calculates an estimation of the value of the stability performance indicator $\hat{\mathcal{K}}_{y}$, considering the application of the $i$-th CCRC.
\end{definition}

\vspace{-0.7cm}
\section{Data-driven MCDM for CCR Assignment}
\label{sec: MODM}
\begin{figure}[b]
\vspace{-0.8cm}
    \centering    \includegraphics[width=0.95\linewidth]{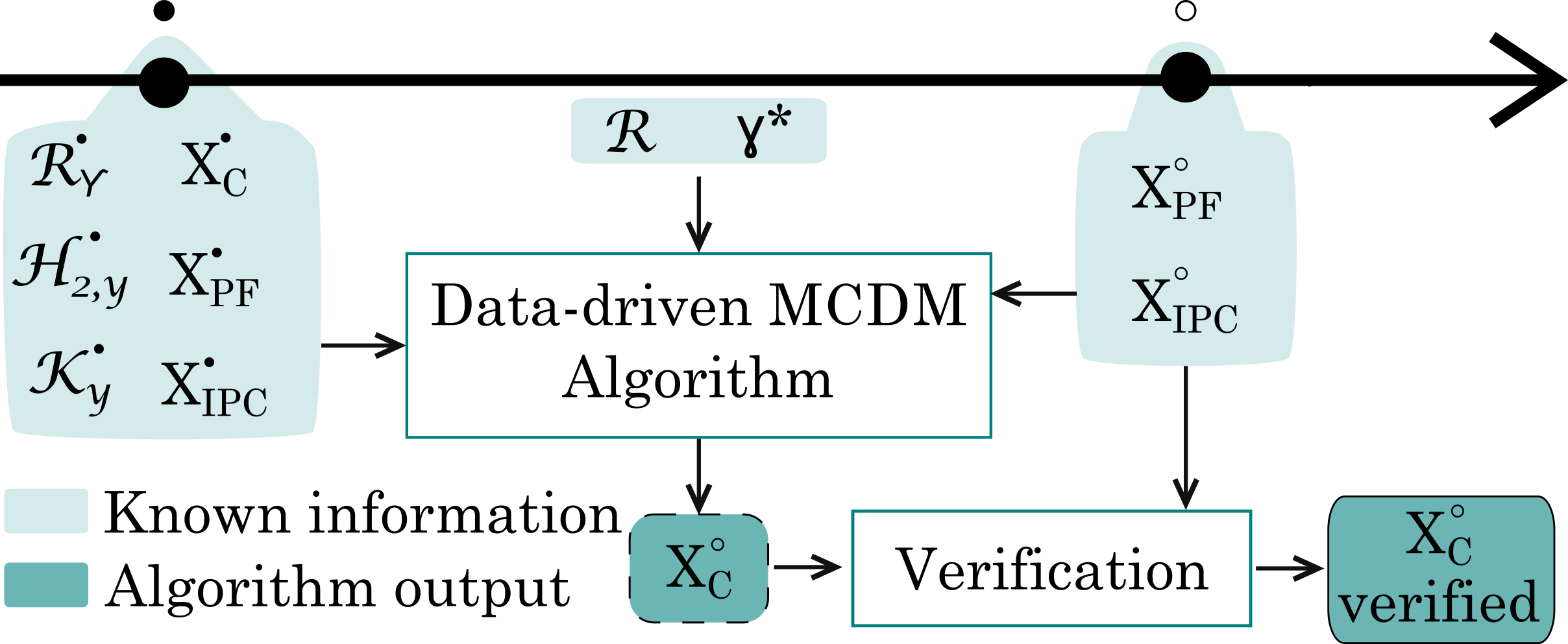}
    \caption{Workflow of the data-driven MCDM algorithm}
    \label{fig: dm_alg}
\end{figure}

\vspace{-0.1cm}
The conventional MCDM problem formulation \cite{triantaphyllou_2000} is adapted for this study, specifically for assigning CCRCs to achieve optimal stability performance within the online timeframe of power system operation. Fig.~\ref{fig: dm_alg} illustrates the MCDM algorithm application, summarizing its main steps along with the input and output quantities. Consider that the system has to transit from one OP, the one marked as a full circle ($\bullet$) in Fig.~\ref{fig: dm_alg}, to another, marked as an empty circle ($\circ$). The MCDM problem consists of finding, among all the possible alternatives, the CCRC to be assigned at $OP^\circ$ (i.e., $X_C^\circ$) that makes the system stable and with the best stability performances
, according to multiple indicators. Concerning the online implementation of the algorithm, this goal can be achieved thanks to the use of data-driven \textit{surrogate} models for the computation of the stability ($\hat{\Upsilon}$) and the stability performance indicators ($\hat{\mathcal{H}}_{2,y}$ and $\hat{\mathcal{K}}_{y}$). However, the speed-up in computation 
is paid for in terms of accuracy. Therefore, before operating the system with the proposed $X_C^\circ$, a verification of the stability response through the \textit{exact} models is carried out. Hence, as depicted in Fig.~\ref{fig: dm_alg}, the main steps of the algorithm are the data-driven MCDM and the verification of the data-driven solution. Next, the problem formulation is described. 
\vspace{-0.4cm}
\subsubsection{Problem Alternatives} 
The decision-making problem \textit{alternatives} are the CCRCs. 
In principle, all the possible CCRCs are the \textit{variations without repetitions} of the CCRs. Their number is therefore equal to the number of different CCRs, raised to the power of the number of IPCs in the system. Many of these CCRCs are not feasible, as does not fulfill the operating principles identified in \cite{gomis2020principles}. However, the number of feasible CCRCs is still high, especially for power systems with high penetration of IPCs with multiterminal configuration. Considering such a large number of CCRCs in stability studies, especially if it is intended to employ data-driven tools, adds complexity. Therefore, it is proposed to select a reduced number of feasible CCRCs, according to a selection criterion explained in Section \ref{sec: CCRCs_selection}.\\
Define this selection of CCRCs as $\mathcal{R}$, representing the full set of \textit{alternatives}. However, the actual set of \textit{alternatives} considered each time the MCDM algorithm is run changes according to two constraints. 
First, it has to be considered that it is possible to assign a CCRC to one OP only if the CCRC is stable during the transition. That is, the proposed CCRC has to be stable both in $OP^\bullet$  and $OP^\circ$. Therefore, if $\mathcal{R}_\Upsilon^{\bullet}$ and $\mathcal{R}_\Upsilon^{\circ}$ collect the CCRCs that result stable at the corresponding OP, the following condition has to be true:
\vspace{-0.3cm}
\begin{equation}
    X_C^{\circ}\in \mathcal{R}_\cap \quad \text{with} \quad \mathcal{R}_\cap=\mathcal{R}_\Upsilon^{\bullet} \cap \mathcal{R}_\Upsilon^{\circ}
    \label{eq: constr1}
\vspace{-0.2cm}
\end{equation}
\noindent The second constraint is related to how many simultaneous CCRs changes are allowed when transiting from $X_C^{\bullet}$ to $X_C^{\circ}$.
Consider that a different CCRC assignment can involve the change of a single CCR or multiple CCRs. In case multiple CCR changes are involved, their switch has to happen simultaneously. As large transients might be caused by the difficult coordination, it might be preferred to constrain the decision-making algorithm to look for the best CCRC which implies a limited number of CCRs changes. Indicating $\gamma$ as the number of CCR changes that the assignation of $X_C$ at $OP^\circ$ involves, such a constraint can be expressed as $\gamma \leq \gamma^*$,
where $\gamma^*$ is the maximum number of allowed CCRs changes. The value of $\gamma^*$ can be defined by the System Operator (SO). The two mentioned constraints can be summarized as
\vspace{-0.2cm}
\begin{equation}
    X_C^{\circ}\in\mathcal{R}'_\cap \quad \text{with} \quad \mathcal{R}'_\cap=\mathcal{R}_\Upsilon^{\bullet}\cap\{\mathcal{R}_\Upsilon^{\circ}|\gamma\leq\gamma^*\}
    \label{eq: merge_constr}
    \vspace{-0.2cm}
\end{equation}
\noindent According to \eqref{eq: merge_constr}, the 
solution of the MCDM problem $X_C^{\circ}$ belongs to the set of \textit{alternatives} $\mathcal{R}'_\cap$. This set collects all the CCRCs that are stable at both the current and successive OPs and involve a number of CCRs changes lower than $\gamma^*$. The value set for $\gamma^*$ might lead to $\mathcal{R}'_\cap=\emptyset$. In this case, it is proposed to relax the constraint 
by increasing $\gamma^*$ until at least one CCRC stable at both OPs is found (i.e. $\mathcal{R}'_\cap\neq\emptyset$).\\
\vspace{-0.3cm}
\subsubsection{Problem Criteria and Performances}
The \textit{criteria} ($\zeta$) used to compare the alternatives are the enhancement in the stability performances, measured as the improvement of the stability performances indicator for each considered signal. 
Denoting by $\mathcal{Y}$ the set of all signals included in the analysis, the number of criteria in the MCDM algorithm is equal to 2$|\mathcal{Y}|$, as both the improvement of the $\mathcal{H}_2$-norm and of the DC Gain  are considered for each signal. Therefore, for a signal $y$, the value of the performance ($\rho$) of the $i$-th \textit{alternative}, according to the $j$-th \textit{criterion} is calculated as in (\ref{eq: perf}a) or (\ref{eq: perf}b), whether it refers to the $\mathcal{H}_2$-norm or the DC Gain.
\vspace{-0.2cm}
\begin{equation}
\rho_{i,j}=\hat{\mathcal{H}}_{2,y}^{\circ}-\mathcal{H}_{2,y}^{\bullet}  
    \quad \text{(a)}\quad;\quad
    \rho_{i,j}=\hat{\mathcal{K}}_{y}^{\circ}-\mathcal{K}_{y}^{\bullet}\quad \text{(b)}
    \label{eq: perf}
    \vspace{-0.2cm}
\end{equation}
\noindent The decision to adopt as criteria the improvement of the indicators values rather than their absolute values stems from the fact that these indicators can sometimes approach zero (i.e. their minimum). In such cases, using weights in a weighted sum becomes ineffective. Conversely, an indicator close to zero indicates a favorable stability condition for the system and should be rewarded accordingly. Therefore, by adopting the formulations in (\ref{eq: perf}), an improvement (of the $i$-th \textit{alternative}, according to the $j$-th \textit{criterion}) is achieved when the value of $\rho_{i,j} < 0$. Hence, the best \textit{alternative} of the problem is the one with the lowest weighted sum of performance terms.\\
In expressions (\ref{eq: perf}), the values of the indicators with the operating conditions in $\bullet$ are calculated with the \textit{exact} models, while the performance indicators at $\circ$ are calculated using the data-driven \textit{surrogate} models. This approach is chosen because the \textit{exact} information about the stability and stability performances for the current OP ($\bullet$) is assumed to be known. Whereas, the stability performance at the subsequent OP ($\circ$), considering the assignments of all the possible CCRC in $\mathcal{R}'_{\cap}$, is estimated via the \textit{surrogate} models to accelerate computation compared to using the \textit{exact} model.
\subsubsection{Complete Performance Matrix and Solution} 
For the problem under consideration, the full MCDM performance matrix takes the form shown in (\ref{eq: perf_matrix_ddmodm}).
\vspace{-0.3cm}
\begin{equation}
\resizebox{0.89\linewidth}{!}{$
\mathbf{\Pi} =\!
\begin{blockarray}{ccc}
& \mathcal{H}_{2,y\in|\mathcal{Y}|} & \mathcal{K}_{y\in|\mathcal{Y}|}\\
 & \overbrace{\zeta_1 \quad \dots \quad \zeta_{|\mathcal{Y}|}} & \overbrace{\zeta_{|\mathcal{Y}|+1} \quad \dots \quad \zeta_{2|\mathcal{Y}|}} \\
\begin{block}{c@{\!\;}[c|c]}
  X_{C_1} & \rho_{1,1} \quad \dots \quad \rho_{1,|\mathcal{Y}|} & \rho_{1,|\mathcal{Y}|+1} \quad \dots \quad \rho_{1,2|\mathcal{Y}|} \\
  \vdots &  \vdots \qquad  \ddots \qquad \qquad &  \vdots \qquad  \ddots \qquad \qquad \\
   X_{C_{|\mathcal{R}'_\cap|}} & \rho_{|\mathcal{R}'_\cap|,1} \qquad \quad \rho_{|\mathcal{R}'_\cap|,|\mathcal{Y}|} & \rho_{|\mathcal{R}'_\cap|,|\mathcal{Y}|+1} \quad \rho_{|\mathcal{R}'_\cap|,2|\mathcal{Y}|} \\
\end{block}
\end{blockarray}
$}
\label{eq: perf_matrix_ddmodm}
\vspace{-0.3cm}
\end{equation}
The $X_C^\circ$ to be assigned at $OP^\circ$, to provide the best small-signal stability performances, is the solution of the data-driven MCDM problem. It is the $i^*$-th CCRC, where $i^*$ is 
\vspace{-0.1cm}
\begin{equation}
    \textstyle i^* = 
    \arg\min_{i} \sum_{j}^{2|\mathcal{Y}|}{w_{\zeta_j} \rho_{i,j}} \quad \text{with} \quad i =1,..., |\mathcal{R}'_\cap|
    \label{eq: objfun}
\vspace{-0.1cm}
\end{equation}
\noindent where $w_{\zeta_j}$ is the weight of the $j$-th criterion.

\subsubsection{Solution Verification}
 The stability of $X_C^{\circ}$ is assessed through the \textit{exact} model $\Psi$. If it is confirmed as a stable solution it is assigned for operating the system. Otherwise, the stability of the others $X_C \in \mathcal{R}'_\cap$ is assessed, starting from the one with better performances. The first CCRC providing a stable system response is considered the MCDM final solution.

\vspace{-0.5cm}
\section{Training The Data-driven Surrogate Models}
\vspace{-0.1cm}
The MCDM algorithm illustrated in Section \ref{sec: MODM} bases the computation of the \textit{performance} matrix on the data-driven \textit{surrogate} models defined in Section \ref{sec: ss_ind}. This Section proposes a methodology for obtaining such \textit{surrogate} models. 
First, the set of CCRCs \textit{alternatives} is selected. Then, for each selected CCRC a training data set is generated by computation, employing the \textit{exact} models. The obtained data sets are properly arranged. Finally, the classification and regression \textit{surrogate} models are trained. Next, the mentioned steps are described.   

\vspace{-0.5cm}
\subsection{Select a Reduced Set of CCRCs Feasible Alternatives}
\label{sec: CCRCs_selection}
\vspace{-0.1cm}
One possibility is considering as the set of possible CCRCs \textit{alternatives} ($\mathcal{R}$) all the feasible CCRCs. That is, the CCRCs fulfilling the operating principle stated in \cite{gomis2020principles}. According to the size and topology of the system, i.e. the number of IPCs in the grid and whether they are involved in multi-terminal connections, the number of feasible CCRCs can rise a lot. This presents an added challenge for data-driven analysis, as data must be generated for each CCRC. To address this, this study proposes a methodology for identifying a reduced set of CCRCs that maintains good stability performance across the entire operating space based on multiple performance indicators. This approach extends the methodology introduced in \cite{10202947}, which was initially developed for a single indicator.\\ 
The methodology introduced in \cite{10202947} starts with generating a data set of random OPs, considering all feasible CCRCs. For each OP the small-signal stability is assessed and the performance indicators are obtained. Then, a clustering process is applied to identify and group the CCRCs that have similar stability performance behavior in the same sub-regions of the operating space. Finally, a Decision Tree (DT) regression algorithm extracts the behavior of the clusters in the form of conditional rules, which are used to select the final set of CCRCs that cover the entire operating space with the best performances.

\begin{figure}[h!]
    \centering
    \subfloat[Proposed sets intersection visualization.]{
    \includegraphics[width=0.5\linewidth]{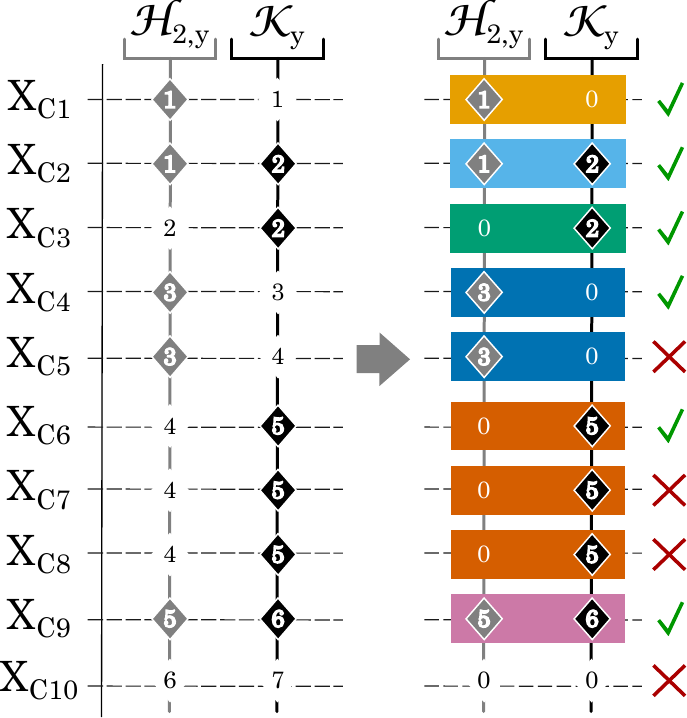}
    \label{fig: set_inters_a}}
    \subfloat[Equivalent Venn diagrams representation.]{
    \includegraphics[width=0.5
\linewidth]{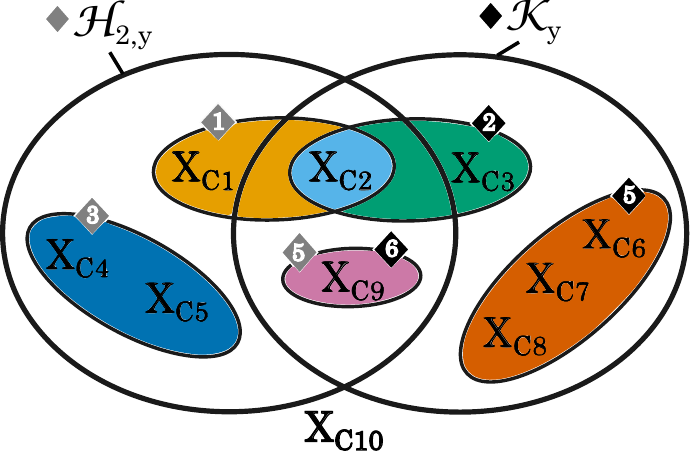}
       \label{fig: set_inters_b}
       }
    \caption{Sets intersection visualization and CCRCs selection.}
    \label{fig: set_inters}
\vspace{-1cm}
\end{figure}

The process described above 
can be repeated for multiple stability performance indicators. 
To achieve the final selection that accounts for multiple indicators, intersecting set visualization techniques are employed. Specifically, the UpSet method proposed in \cite{6876017}, designed to visualize interactions among more than three sets, is adapted to this context. 
Fig.~\ref{fig: set_inters} illustrates this visualization method. The CCRCs to be grouped are analyzed based on their attributes, defined as follows: 
\begin{itemize}
    \item \textit{Attribute 1}: Cluster membership. Each CCRC is associated with a cluster, indicated in Fig.~\ref{fig: set_inters} by a number (e.g., 1 = Cluster 1, 2 = Cluster 2, etc.).
    \item \textit{Attribute 2}: Whether the cluster to which a CCRC belongs is selected. This is shown in Fig.~\ref{fig: set_inters} using a rhomboidal mark (\raisebox{-0.2em}{\includegraphics[height=1em]{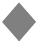}} for selection based on $\mathcal{H}_{2,y}$, and \raisebox{-0.2em}{\includegraphics[height=1em]{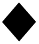}} for selection based on $\mathcal{K}_{y}$).
\end{itemize}
Since cluster membership is irrelevant for CCRCs in unselected clusters, these CCRCs are first grouped into a generic Cluster 0. Next, a combined attribute is created to describe CCRCs, capturing the combination of cluster memberships for selected clusters. In the example shown in Fig.~\ref{fig: set_inters_a}, seven such combinations are identified (\raisebox{-0.2em}{\includegraphics[height=1em]{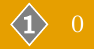}}, \raisebox{-0.2em}{\includegraphics[height=1em]{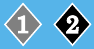}}, \raisebox{-0.2em}{\includegraphics[height=1em]{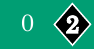}}, etc.). CCRCs with the same combined attribute are then grouped via a set intersection process. Finally, one CCRC is selected from each group (e.g., the first one in the list) and added to the set of selected CCRCs, $\mathcal{R}$. In this way, $\mathcal{R}$ collects the CCRCs required to operate in the entire system operating space with enhanced small-signal stability performances, according to multiple indicators. For completeness, Fig.~\ref{fig: set_inters_b} offers an equivalent representation of Fig.~\ref{fig: set_inters_a} using Venn diagrams. Although this familiar technique effectively illustrates the same concept, it becomes less readable when dealing with many elements and intersections across multiple indicators. In such cases, the visualization method proposed in Fig.~\ref{fig: set_inters_a} is more suitable.
\vspace{-0.3cm}
\subsection{Data Generation}
\label{sec: datagen}
To train and validate the data-driven \textit{surrogate} models, data are required. In particular, it is necessary to collect information about the small-signal stability and stability performances in the operating space, considering the assignment of each CCRC in $\mathcal{R}$. Since these scenarios are not available during the system’s real operation, it is necessary to generate such data by computation. The computing tools involved are the PF calculation and the \textit{exact} models $\Psi$, $\Gamma_{\mathcal{H}_{2,y}}$, and $\Gamma_{\mathcal{K}_{y}}$ $\forall y \in \mathcal{Y}$. For each CCRC $\in \mathcal{R}$ the data generation process is carried out twice, once for generating the training data and once for generating data for final validation. Two different sampling strategies are applied. To generate training data, the method introduced in \cite{rossi2022data} is utilized. This approach focuses on efficiently producing high-quality data for training stability predictive models. It enables comprehensive exploration of the system's stability across its entire operating space while ensuring higher granularity within the stability margin. The process employs Latin Hypercube Sampling (LHS) to randomly select OPs and uses the Entropy function to identify the stability margin and guide the sampling process toward it. To generate validation data, $N_{OP}$ random OPs are sampled in the entire system operating space by the use of LHS. After the data are generated, they are organized into training and test data sets as indicated in Definitions \ref{def: dd_stab}-\ref{def: dd_dcgain}.
\vspace{-0.5cm}
\subsection{Features Engineering}
Feature engineering is applied to preprocess data in the training and test data sets. This process consists of creating, transforming, and selecting relevant features from raw data to improve the performance of ML models. It involves techniques such as data cleaning, scaling, encoding categorical data, handling outliers, and creating new features that better represent the underlying patterns in the data.


\subsubsection{Creating New Features}
Additional input features are created starting from the variables collected in $X_{PF}$ and $X_{IPC}$. Specifically, the added features are:
\vspace{-0.1cm}
\begin{itemize}
    \item The apparent power injected by generators, at both sides of the IPCs, absorbed by the loads, and injected or absorbed by the Thevenin's equivalents. 
    \item Direction of the PF: This applies to system assets such as IPCs and Thevenin equivalents, which can experience PF in both directions. 
    Boolean features indicating the sign of the power flow can be 
    added to the datasets.
    \vspace{-0.1cm}
\end{itemize}
\subsubsection{Data Cleaning}
The data cleaning process includes removing features containing a single value, duplicated features, and correlated features. To identify correlated features, the Pearson correlation coefficient is used. The coefficient is computed for each pair of variables, and pairs with a high coefficient 
are flagged for review to ensure critical system information is preserved. For highly correlated pairs, one variable is considered for removal based on the following criteria:  
\begin{itemize}
    \item Power-related variables ($P$, $Q$, $S$): Both 
    retained.  
    \item Features associated with different nodes: Both 
    retained.  
    \item Other cases: The feature with the lower priority is removed, following this hierarchy: currents, AC voltages, DC voltages, and angles. 
\end{itemize}
\vspace{-0.1cm}
\subsubsection{Data Scaling}
The input quantities and, for the regression case, the output data are scaled. 
\subsubsection{Handling Outliers} 
Winsorizing is considered, which replaces the outliers with the values of a specified percentile.
\vspace{-0.5cm}
\subsection{Models Training}
As outlined in Definitions \ref{def: dd_stab}-\ref{def: dd_dcgain}, the data-driven \textit{surrogate} models designed to substitute the \textit{exact} models consist of a classifier to predict system stability and regression models to estimate the values of the stability performance indicators. Various classification and regression algorithms could effectively perform these tasks. To determine the most suitable ML technique, a comparison of their performance is conducted by training and testing multiple algorithms on the generated data. First, appropriate metrics are chosen to evaluate the performance of the algorithms. Next, several algorithms are trained, and their accuracy is assessed. Based on this comparison, either the best-performing algorithm is selected or the top-performing algorithms are shortlisted as candidates. Subsequently, the training of the selected model(s) is refined through feature selection and hyperparameter tuning. A detailed explanation of these steps is provided next.

\subsubsection{Set Metrics}
The classification task involves predicting the system's stability. Misclassifying an unstable case as stable poses a critical risk to system operation, whereas misclassifying a stable case as unstable, while inconvenient, is less severe. Thus, a conservative approach prioritizes accurately predicting true stable cases. Precision and Recall are key metrics for evaluating the model's performance in this context. Precision 
measures the proportion of true stable cases out of all cases predicted as stable (i.e., true stable and false stable) \cite{sammut2011encyclopedia}. Recall 
evaluates the proportion of true stable cases out of all cases that are actually stable (i.e., true stable and false unstable) \cite{sammut2011encyclopedia}.
A metric that combines Precision and Recall can be useful for this task. The $F_{\beta}$ score \cite{sammut2011encyclopedia} 
is defined as the harmonic mean of Precision and Recall, weighted by a factor $\beta$. Typically, $\beta=0.5$ or $\beta=2.0$ is chosen, depending on whether Precision or Recall should be given more weight, respectively. The balance of instances in the training data significantly affects model training, as the model is more likely to predict the dominant class. Hence, this work proposes adopting $\beta=2.0$ when less than 50\% of the input data represents stable cases, prioritizing Recall to minimize false unstable. Conversely, if more than 50\% of the data is stable, $\beta=0.5$ is used to prioritize Precision and reduce false stable.


Concerning the regression models fitting the stability performance indicators, the metrics proposed is the $R^2$ score.
    
\subsubsection{Compare Models Performances}
To determine the most suitable ML model for each \textit{surrogate} model, a performance-based comparison is conducted across all options. 
A k-Fold Cross-Validation (kFCV) approach is employed, and the comparison is based on the selected evaluation metrics. 



\subsubsection{Permutation Feature Importance}
Once the best-performing algorithm or the top-performing algorithms are identified, 
a feature selection can be performed, based on permutation feature importance (PFI).  

\subsubsection{Hyperparameters Tuning}
A Grid-Search kFCV (GSkFCV) process is utilized to tune the hyperparameters of the selected models.

\vspace{-0.8cm}
\section{Case Study}
The proposed data-driven MCDM algorithm is demonstrated using a hybrid AC/DC test system. This section presents a detailed description of the test system, outlines the implementation of the methodology for training the \textit{surrogate} models, provides further insights into the behavior of the stability performance indicators, and showcases the results of the MCDM algorithm implementation. 

\vspace{-0.6cm}
\subsection{Test System Description}
\label{sec: test_system}
\begin{figure}[h]
    \centering
    \includegraphics[width=0.8\linewidth]{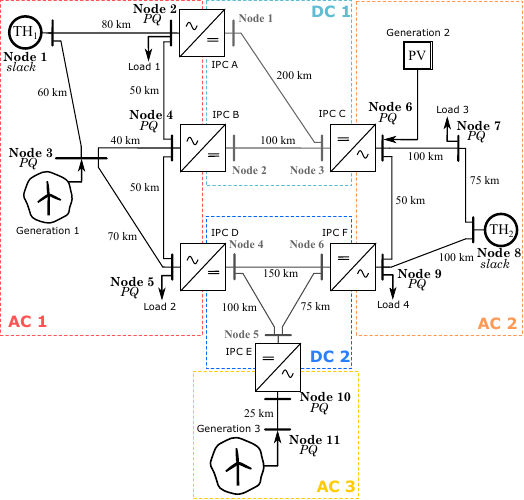}
    \caption{Hybrid AC/DC grid test system \cite{10017135}.}
    \label{fig: power_syst}
    \vspace{-0.3cm}
\end{figure}

\vspace{-0.2cm}
The test system used throughout this study is the hybrid AC/DC grid represented in Fig.~\ref{fig: power_syst} \cite{10017135}. 
This grid consists of three AC and two HVDC sub-grids, interconnected by six IPCs. 
In both AC-1 and AC-2 sub-grids there are:
\begin{itemize}
    \item Two loads, representing the aggregated power demand.
    \item One renewable-based generator, representing, in aggregated form, wind or photovoltaic-based power generation.
    \item A Thevenin equivalent for the rest of the grid.
\end{itemize}
The sub-grid AC-3 has one wind-based power plant, which might represent a large off-shore wind farm.
\vspace{-0.2cm}

The system's operating space is defined by the operating ranges of the generators and the minimum and maximum power demand. Table \ref{tab: gen_range} summarizes the operating ranges of the generators, including the minimum and maximum active power they can inject, set to 5\% and 95\% of their nominal power, respectively. Regarding reactive power, all generators are assumed to support power factors ranging between 0.8 and 0.95. 
Table \ref{tab: dem_range} outlines the load consumption ranges, where the total power demand varies between 200 MW and 700 MW. Individual loads are assumed to contribute to the total demand within a range of $\pm$30\% of their base load, defined as a percentage of the total demand.
\vspace{-0.3cm}
\begin{table}[h]
\caption{Generators operating range.}
\label{tab: gen_range}
\resizebox{\columnwidth}{!}{%
\begin{tabular}{ccc|ccc}
\toprule
 & Minimum & Maximum &  & Minimum & Maximum \\ \midrule
$P_{G_1}$ {[}MW{]} & 15 & 285 & $\cos \phi_{G_1}$ & 0.8 & 0.95 \\
$P_{G_2}$ {[}MW{]} & 5 & 95 & $\cos \phi_{G_2}$ & 0.8 & 0.95 \\
$P_{G_3}$ {[}MW{]} & 7.5 & 142.5 & $\cos \phi_{G_3}$ & 0.8 & 0.95 \\ \bottomrule
\end{tabular}%
}
\vspace{-0.3cm}
\end{table}

\begin{table}[t]
\centering
\caption{Power demand range}
\label{tab: dem_range}
\begin{tabular}{cccc}
\toprule
 & Base Load & Minimum & Maximum \\ \midrule
$P_D$ {[}MW{]} & {[}-{]} & 200 & 700 \\
$P_{L_1}$ & 30\% $P_D$ & 21\% $P_D$ & 39\% $P_D$ \\
$P_{L_2}$ & 20\% $P_D$ & 14\% $P_D$ & 26\% $P_D$ \\
$P_{L_3}$ & 20\% $P_D$ & 14\% $P_D$ & 26\% $P_D$ \\
$P_{L_4}$ & 30\% $P_D$ & 21\% $P_D$ & 39\% $P_D$\\
\bottomrule
\end{tabular}
\vspace{-0.7cm}
\end{table}

The IPCs interconnecting the AC and DC grids are MMCs. Each IPC is provided with three control roles. Following the definitions provided in \cite{10443561}, such control roles are:
\begin{itemize}
    \item AC-GFM control that controls the voltage on the AC side 
    \item DC-GFM control that controls the voltage on the DC side through a droop control.
    \item GFL control that doesn't control the voltage of any terminal and synchronizes with the AC grid by a PLL. 
\end{itemize}
\vspace{-0.1cm}
\noindent Given the possibility of implementing such CCRs in each IPC, the number of possible CCRCs is theoretically $N_{CCRCs}=(N_{CCRs})^{|\mathcal{C}|}=3^6=729$, where $N_{CCRs}$ is the number of CCRs per IPC, and $|\mathcal{C}|$ the number of IPCs in the system. However, a large number of these CCRCs result unfeasible and have to be discarded, following the operating principles identified in \cite{gomis2020principles}. Such principles of operation state that, in a hybrid AC/DC grid, each sub-grid needs at least one grid-forming unit. This means that during system operation: (i) In each AC sub-grid, at least one node has to be an AC-GFM unit, either an IPC or another type of unit; (ii) In each DC sub-grid, at least one node has to be an IPC operating with DC-GFM control. 
For the test system, it follows that IPC-E can only be an AC-GFM unit and also many other CCRCs must be discarded. Therefore, the number of feasible CCRCs is reduced to $N_{CCRCs}=95$. 
\vspace{-0.6cm}
\subsection{Small-signal Stability Performances}
\vspace{-0.1cm}
The assessment of small-signal stability performance necessitates the analysis of signals capable of capturing the dynamic behavior of both AC and DC grids. Accordingly, this study evaluates frequency signals to characterize AC grid dynamics and voltage signals to capture DC grid dynamics. For the system case study, three frequency signals are analyzed — one per AC grid — along with six voltage signals, related to each DC grid bus. Although stability performance indicators can be computed for each signal individually, it is proposed to evaluate them in a way that provides an overall assessment of the dynamic behavior across all AC and DC grids. Therefore, the transfer functions matrix required for the calculation of the $\mathcal{H}_2$ norms and DC gain (as defined in Definitions \ref{def: exact_h2} and \ref{def: exact_dcgain}) are determined by setting the active and reactive power at all buses as inputs. Regarding the outputs, the transfer functions matrix for frequency uses the frequencies of the three AC grids, while the transfer functions matrix for voltage considers the voltages at all six DC buses. Then, the calculation of the $\mathcal{H}_2$-norm directly provides a single-value indicator that accounts for the total energy of all the signals. In contrast, the DC gain calculation returns a matrix of dimensions \(\mathbb{R}^{n_{\text{outputs}} \times n_{\text{inputs}}}\), which collects the DC gains for each input-output combination. To derive a single-value indicator, the maximum DC gain is considered, representing the highest deviation across all signals. As a result, four stability performance indicators are considered: \( \mathcal{H}_{2,f} \), \( \mathcal{H}_{2,V_{DC}} \), \( \mathcal{K}_f \), and \( \mathcal{K}_{V_{DC}} \).
\vspace{-0.5cm}
\subsection{Reduced Set of CCRCs}
\begin{figure*}[t]
    \centering
    \includegraphics[width=0.9\linewidth]{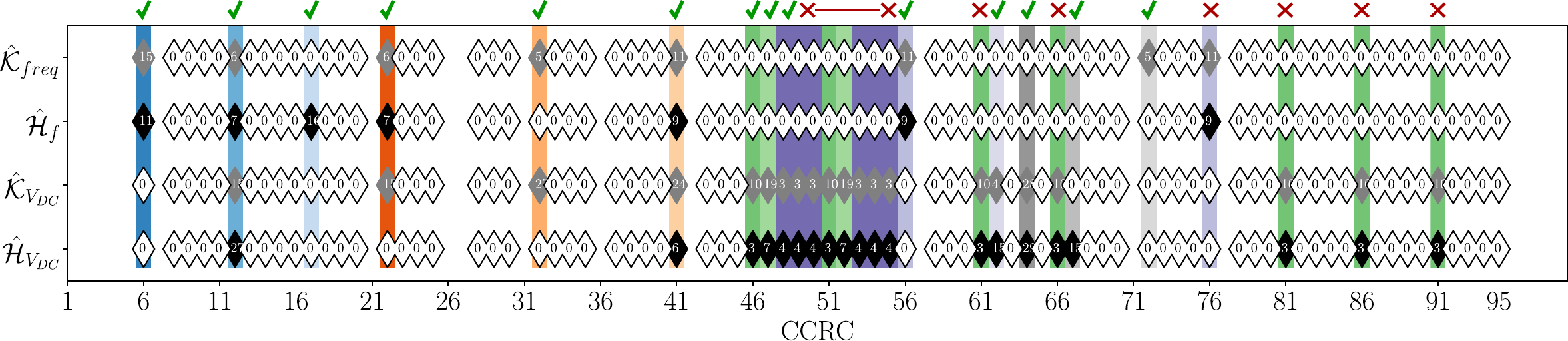}
    \caption{CCRCs selection trough sets intersection visualization.}
    \label{fig: venn}
    \vspace{-0.6cm}
\end{figure*}
\begin{figure}[htbp]
	\centering
    \includegraphics[width=\linewidth]{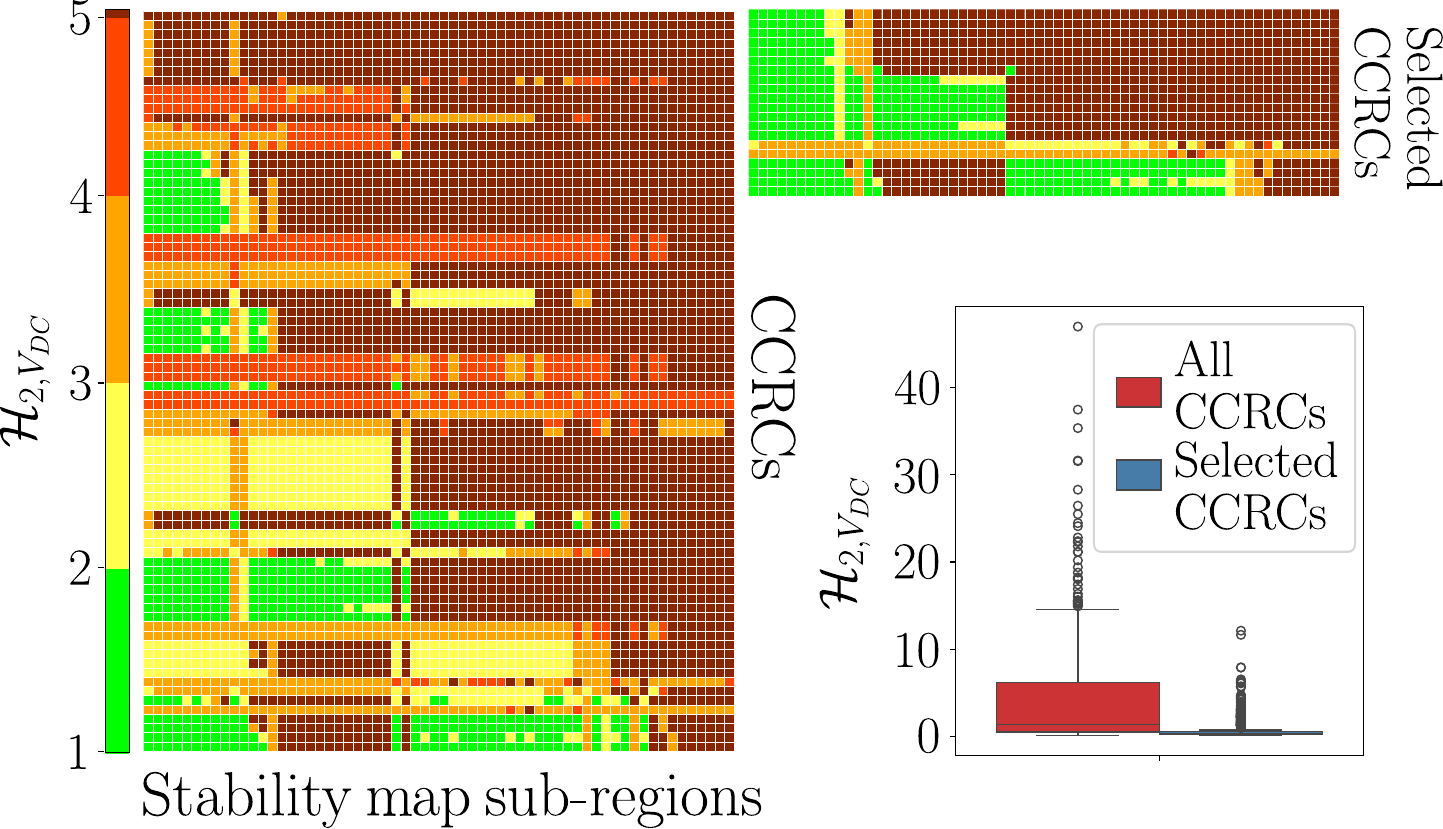}

    \caption{Stability maps considering all CCRCs (left) and a reduced set with better performance (top right) based on $\mathcal{H}_{2,V_{DC}}$. The bottom right panel presents a box plot comparing the distribution of the indicator.}
    \label{fig: stab_maps}
    \vspace{-0.3cm}
\end{figure}

Through the application of the methodology described in Section~\ref{sec: CCRCs_selection}, the minimum set of CCRCs required to operate the system with the best stability performance is identified. The stability behavior of the system across its entire operating space is evaluated for each stability indicator, considering all 95 feasible CCRCs. A clustering algorithm is applied to group CCRCs based on their ability to provide similar stability performance within specific subregions of the operating space. The clustering results are illustrated using stability performance maps, as proposed in \cite{10202947}. Fig.~\ref{fig: stab_maps} displays the map for the $\mathcal{H}_{2,V_{DC}}$ indicator. Maps corresponding to other indicators are omitted for brevity. In this map, each row corresponds to a CCRC, while each column represents a subregion of the 
operating space. 
The color in each cell indicates the stability performance level of a given CCRC in a specific subregion. CCRCs are grouped according to their clusters and ordered from top to bottom, reflecting their average performance across the entire operating space, from worst to best. The top rows in the maps represent CCRCs which consistently exhibit instability across all subregions (performance level = 5). These CCRCs are excluded in successive analyses. The subregions are arranged from left to right, starting with those where CCRCs demonstrate the best average performance and ending with those where performance is poorest.

Following the selection process outlined in \cite{10202947}, a reduced set of CCRCs is identified. Fig.~\ref{fig: stab_maps} presents the stability map for this reduced set, derived by extracting the stability maps of the selected CCRCs from Fig.~\ref{fig: stab_maps}. The figure shows how the chosen CCRCs ensure stable operation across the entire operating space, with the best possible performance. Additionally, the box plot in Fig.~\ref{fig: stab_maps} compares the distribution of the $\mathcal{H}_{2,V_{DC}}$ indicator when using all CCRCs versus the selected subset, highlighting a significant reduction in its values.

The CCRC clustering and selection process is also applied to the $\mathcal{H}_{2,f}$, $\mathcal{K}_{V_{DC}}$, and $\mathcal{K}_{f}$ indicators. As outlined in Section~\ref{sec: CCRCs_selection}, the clustering results are visualized using a set representation, as shown in Fig.~\ref{fig: venn}. This representation facilitates the final selection of the minimum number of CCRCs required to ensure good stability performance across the entire operating space, considering multiple indicators. The number of CCRCs is minimized by retaining only one from each group of CCRCs that exhibit similar stability performance within the same regions of the operating space. 

\subsubsection*{Insight on Indicators Behavior}
\begin{figure}
    \centering
    \includegraphics[width=0.9\linewidth]{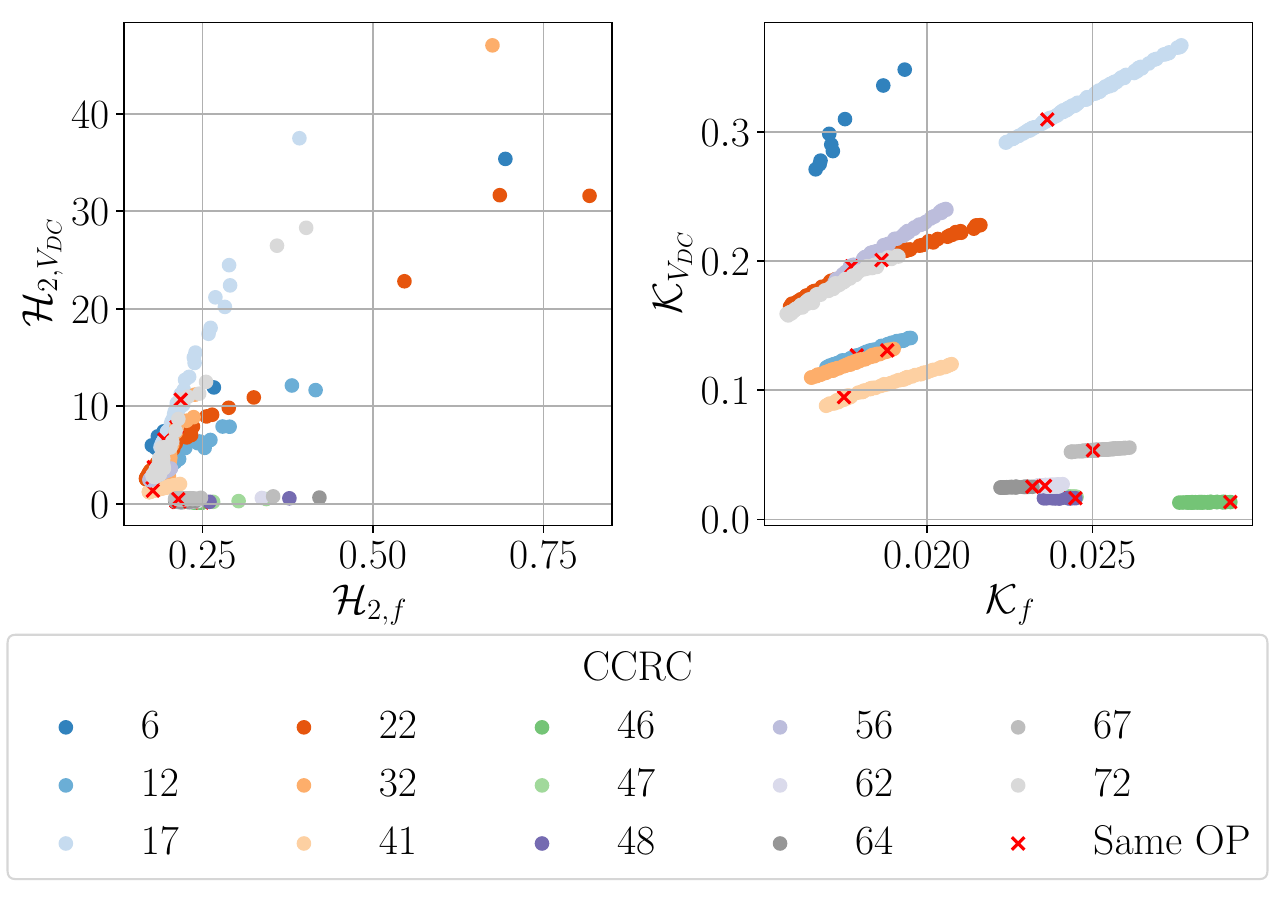}
    \caption{Relations and trends of indicators.}
    \label{fig: ind_behav}
\vspace{-0.7cm}
\end{figure}

\begin{figure}[h!]
    \centering
    \includegraphics[width=0.9\linewidth]{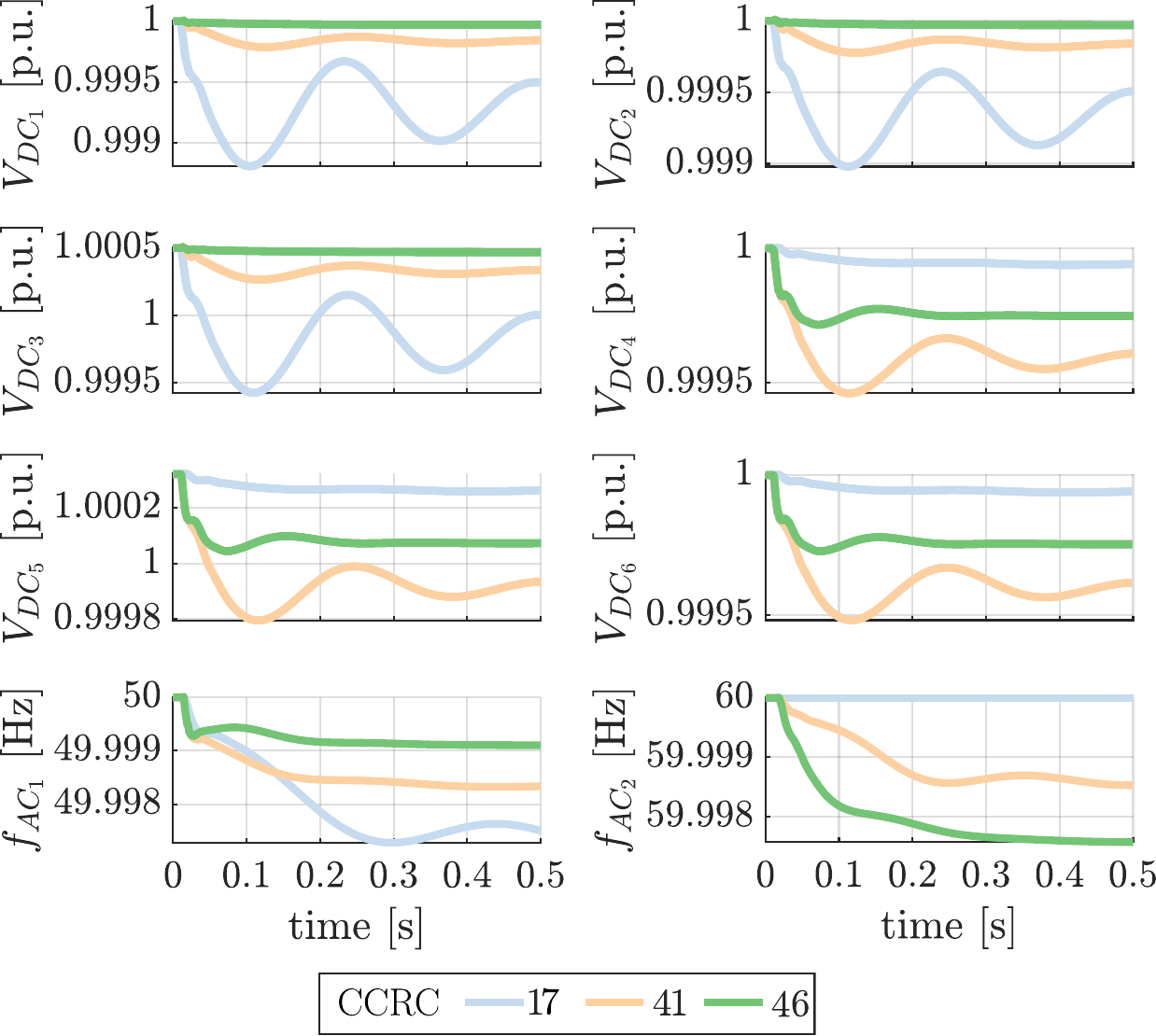}
    \caption{DC voltages and AC frequencies comparison for the same OP, considering the application of CCRCs providing best and worst performances.}
    \label{fig: npf50}
\vspace{-0.5cm}
\end{figure}

Fig.~\ref{fig: ind_behav} illustrates the trends of the stability indicators by comparing their values across 100 random OPs, considering only the stable cases. Specifically, \( \mathcal{H}_{2,f} \) is plotted against the \( \mathcal{H}_{2,V_{DC}} \), while \( \mathcal{K}_{f} \) is compared with the \( \mathcal{K}_{V_{DC}} \). The results indicate that, for all CCRCs, the indicators related to DC voltage and frequency exhibit a positive correlation. This suggests that, when a single CCRC is applied, operating points with improved frequency dynamics also tend to display better DC voltage dynamics, and vice versa. However, when comparing the same operating points under different CCRCs, a trade-off emerges: CCRCs that enhance frequency dynamics tend to deteriorate DC voltage dynamics. This effect is particularly evident when selecting a single operating point and analyzing its indicator values across all CCRCs for which it remains stable. For the selected operating point (marked with a red cross in Fig.~\ref{fig: ind_behav}), applying CCRC number 46 
yields the best DC voltage dynamics but the worst frequency dynamics. Conversely, to achieve optimal frequency dynamics, the assigned CCRC should be CCRC number 41. 
Fig.~\ref{fig: npf50} compares the DC bus voltage dynamics and AC grid frequency dynamics for the selected operating point, considering the application of CCRCs number 17, 41, 46. 
A 1\% load increase at Load 1 is applied as a disturbance. The plots confirm the insights provided by the stability indicators, demonstrating that CCRC number 46 
achieves the smallest maximum DC voltage deviation. However, this CCRC does not yield good frequency dynamics, which are instead better with CCRC number 41. 
Lastly, the dynamics resulting from CCRC number 17 
are included to illustrate the worst-case performance scenario in terms of DC voltage deviation.
\vspace{-0.5cm}
\subsection{Generated Data Sets}
\vspace{-0.1cm}
The data generation process is carried out considering the application of the 14 CCRCs in $\mathcal{R}$. The data set for training $\hat{\Psi}$, $\mathcal{D}_\Upsilon$, collects 11.421 instances, with 46\% of stable cases. 
\vspace{-0.6cm}
\subsection{Models Training}
\subsubsection{Training of $\hat{\Psi}$}
Given the composition of the data set $\mathcal{D}_\Upsilon$, the metrics used to compare and assess the quality of the trained model is the score $F_{\beta = 2}$. An initial model comparison is performed, including a Dummy classifier that predicts a constant value (\( \hat{\Upsilon_i} = 1 \)). The other models considered in the comparison are DT Classifier (DTC), Neural Network (NN) based on Multi-Layer Perceptron (MLP), Extreme Gradient Boosting DTC (XGB-DTC), and Logistic Regression. All algorithms, except XGB-DTC, are implemented using the \textit{Scikit-Learn} library
. The XGB-DTCs model is implemented using the library \textit{XGBoost}. 
For this comparison, all algorithms are tested with their parameters set to default values. It emerges that the MLP and XGB achieve the highest $F_\beta$ score 
with comparable performances. Therefore, both are selected as candidates for training the surrogate model \( \hat{\Psi} \) and subjected to a refined training process. This process includes Features Selection, using PFI, and hyperparameter tuning by GSkFCV. For the MLP model, the tuning process involves testing different activation functions (\textit{relu}, \textit{logistic}, and \textit{tanh}) as well as varying the number of hidden layers and the number of neurons within each layer. For the XGB-DTC model, different values are tested for the learning rate (to control the degree of shrinkage applied to feature weights), the maximum tree depth, and the subsample ratio. Table~\ref{tab: mlp_vs_xgb} summarizes the key features of the models that achieved the highest \( F_\beta \) scores in the GSkFCV and their corresponding performance metrics. Both models achieve a high F-beta score, with only a minimal difference between them. The data-driven MCDM implementation will be demonstrated using the XGB-DTC, as it performs slightly better.

\begin{table}[]
\caption{Best models hyperparameters and scores.}
\label{tab: mlp_vs_xgb}
\begin{tabular}{ccc}
\toprule
Model & \begin{tabular}[c]{@{}c@{}}GSkFCV\\ Hyperparameters\end{tabular} & $F_\beta$ \\ \midrule
MLP & \begin{tabular}[c]{@{}c@{}}Activation function: \textit{relu}\\ 1 hidden layer with 87 neurons\\ ( = number of inputs)\end{tabular} & 0.9758 $\pm$ 0.0027  \\ \midrule
XGB-DTC & \begin{tabular}[c]{@{}c@{}}Learning rate = 0.3\\ Maximum depth of trees = 6\\ Subsamples ratio= 100\%\end{tabular}  & 0.9769 $\pm$ 0.0037 \\ \bottomrule
\end{tabular}
\vspace{-0.2cm}
\end{table}

\subsubsection{Training of $\hat{\Gamma}$}

\begin{figure}
    \centering
    \includegraphics[width=0.9\linewidth]{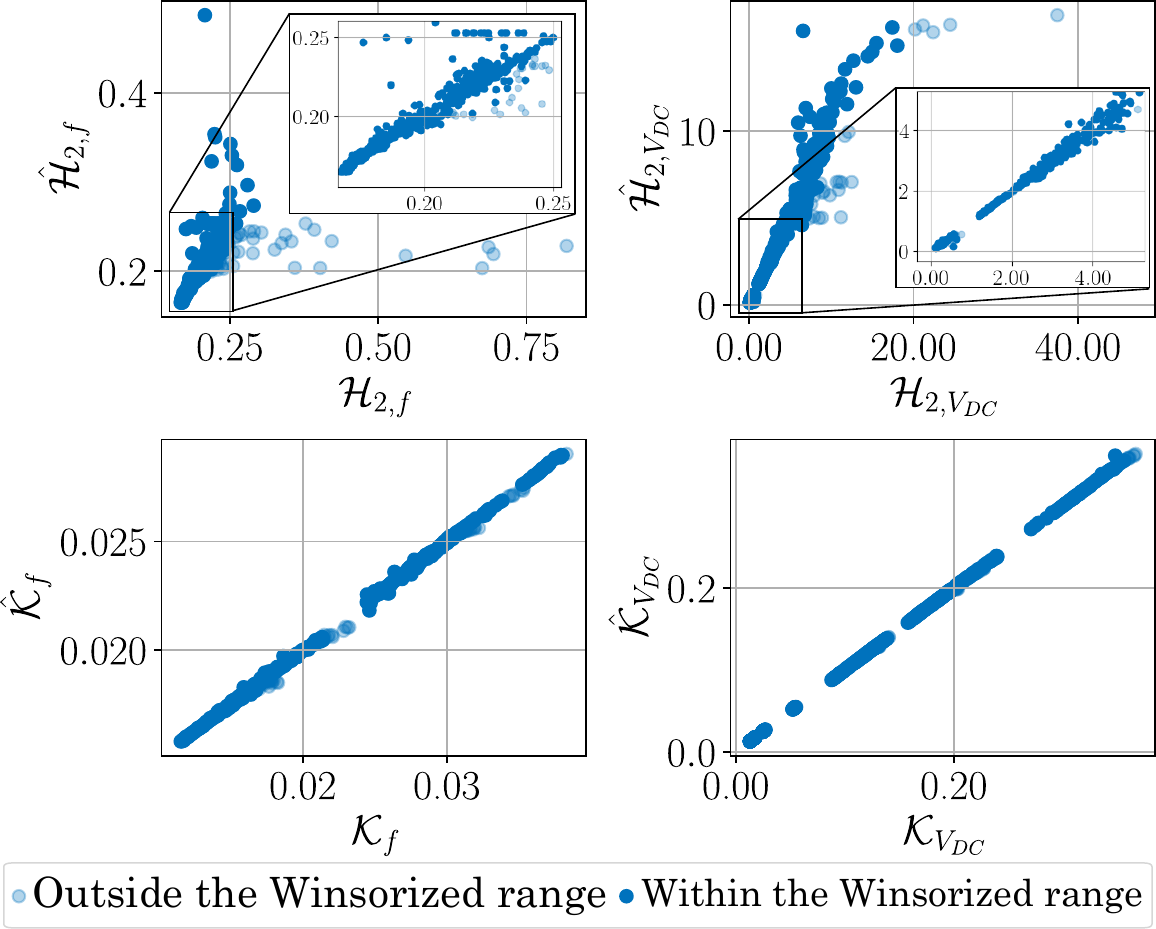}
    \caption{Predicted vs. exact indicators values.}
    \label{fig: ind_acc_zoom}
    \vspace{-0.6cm}
\end{figure}

To compute the stability performance indicators, a separate regression model is trained for each combination of indicator type (\(\mathcal{H}_{2}\)-norm or \(\mathcal{K}\)), signal (\(y \in \{f, V_{DC}\}\)), and CCRC (\(\in \mathcal{R}\)). This approach ensures that each regression model is specifically tailored to its target variable, maximizing accuracy. Consequently, each model is trained independently using the most suitable regression technique among Dummy regressor (that predicts the average value of the target in the training data), Linear Regression, Ridge Regression, DT Regressor (DTR), Neural Network (NN) based on MLP, 
Extreme Gradient Boosting DTR (XGB-DTR). 
PFI is applied only when it enhances the model accuracy. During the training process, 
values are winsorized to the 95th percentile. Model accuracy is evaluated both individually (omitted here for brevity) and in an aggregated manner based on their respective tasks, i.e., models predicting the same indicator for all CCRCs. In the latter case, results are summarized in Table~\ref{tab: rgr_r2}, showing that models predicting the \(\mathcal{K}\) indicators achieve very high accuracy, while those predicting the \(\mathcal{H}_2\)-norm indicators show improved accuracy when points out from Winsorization range are neglected. Fig.~\ref{fig: ind_acc_zoom} compares the predicted and actual indicator values for a test data set, highlighting in the zoomed regions that the regression models for \(\mathcal{H}_2\)-norm indicators achieve better accuracy for lower values. This is particularly relevant, as these lower ranges are of highest interest for obtaining reliable predictions during the execution of the data-driven MCDM algorithm.

\begin{table}[]
\caption{Regression models scores.}
\label{tab: rgr_r2}
\begin{tabular}{ccccc}
\toprule
 & $\mathcal{H}_{2,f}$ & $\mathcal{H}_{2,V_{DC}}$ & $\mathcal{K}_{f}$ & $\mathcal{K}_{V_{DC}}$ \\ \midrule
$R^2$ & 0.3357 & 0.6666 & 0.9993 & 0.9998 \\
\begin{tabular}[c]{@{}c@{}}$R^2$ on\\ winsorized range\end{tabular} & 0.7022 & 0.9906 & 0.9998 & 0.9999 \\ \bottomrule
\end{tabular}
\vspace{-0.3cm}
\end{table}

\vspace{-0.5cm}
\subsection{Data-driven MCDM Algorithm Implementation}
This section demonstrates the implementation of the data-driven MCDM algorithm in a simulated real-world application. To achieve this, two operating scenarios are considered: one based on the day-ahead operation forecast and the other reflecting intra-day forecast updates that approximate actual system conditions. 
Using the day-ahead forecast, the system operator can determine a CCRC assignment schedule by running the MCDM algorithm with exact models, as sufficient computational time is available. Once the intra-day forecast is available, the operator can then re-schedule the CCRC assignment using the proposed data-driven MCDM algorithm.  

\begin{figure}
    \centering
    \includegraphics[width=\linewidth]{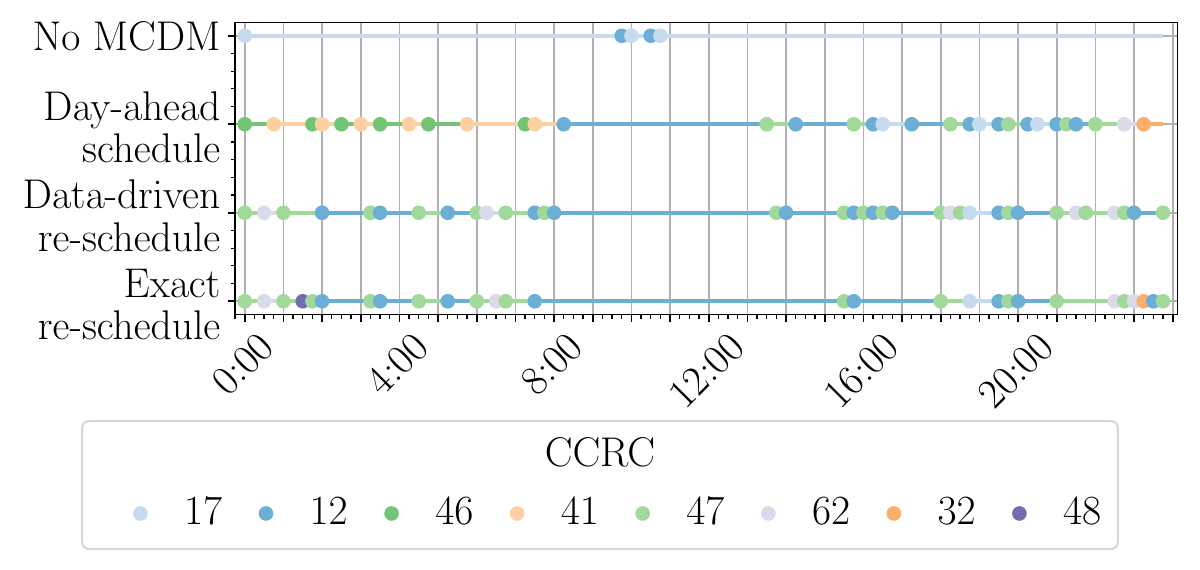}
    \caption{CCRCs assignment schedules comparison.}
    \label{fig: CCRC_schedules}
\vspace{-0.5cm}
\end{figure}

\begin{figure}
    \centering
    \includegraphics[width=\linewidth]{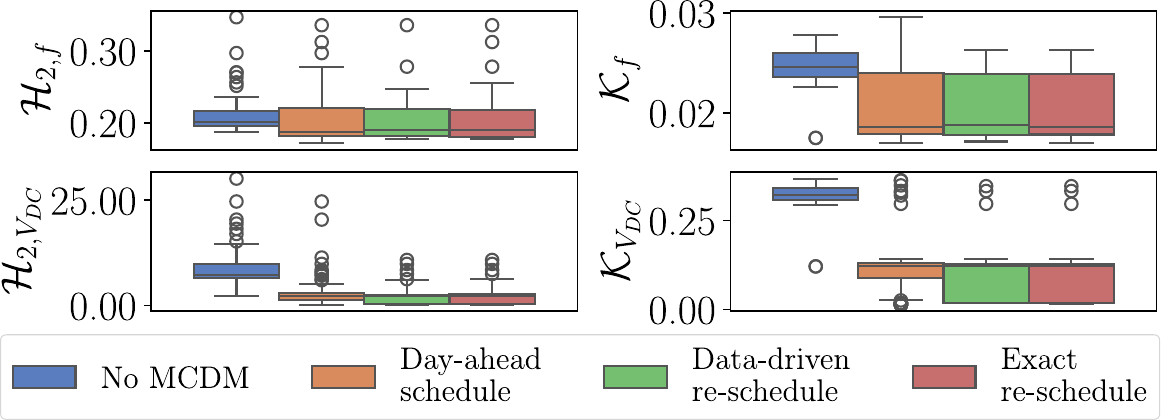}
    \caption{Distributions of the indicators values obtained with the different CCRCs assignment schedules.}
    \label{fig: methods_ind_boxplot}
    \vspace{-0.3cm}
\end{figure}

The effectiveness of the data-driven re-scheduling approach is assessed by comparing its accuracy to the results that would be obtained if the \textit{exact} models were used in the MCDM algorithm for the same operating points. Additionally, the improvement in stability performance achieved through dynamic CCRC re-scheduling is evaluated against the alternative of maintaining the day-ahead CCRC assignment. Finally, an additional scenario is considered in which no MCDM algorithm is used, and CCRC changes are minimized to the extent necessary to ensure system stability. This comparison highlights the stability benefits of a dynamic CCRC assignment. The sequence of CCRC assignments obtained using each scheduling method is compared in Fig.~\ref{fig: CCRC_schedules}, while Fig.~\ref{fig: methods_ind_boxplot} compares their performance based on stability indicator values.

The \textit{Exact re-schedule} serves as the target sequence of CCRC assignments to be achieved. The \textit{No MCDM} approach ensures stable operating points in 100\% of the cases by using only two CCRCs, among the ones that show the highest percentage of stable cases in the training data set. However, the \textit{No MCDM} approach matches the target CCRC assignments only in 5.2\% and leads to very poor stability performance, as shown by the stability indicator distribution in Fig.~\ref{fig: methods_ind_boxplot}. The \textit{Day-ahead re-schedule} matches the target sequence in 43.7\% of cases. However, for the remaining operating points, it does not always guarantee stability, resulting in instability in 14.5\% of cases. In the box plot distribution of Fig.~\ref{fig: methods_ind_boxplot}, these unstable cases have been excluded, yet the performance still remains worse than that of the \textit{Data-driven re-schedule}. The \textit{Data-driven re-schedule} aligns with the target sequence in 89.6\% of cases while also achieving similar stability performance, making it the most effective approach among the alternatives.

\begin{figure}
    \centering
    \includegraphics[width=\linewidth]{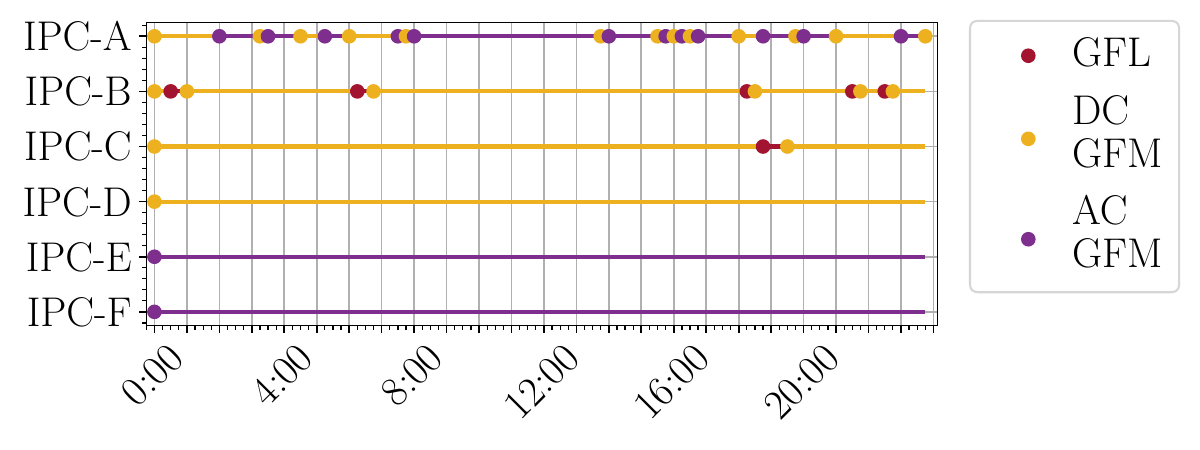}
    \caption{IPCs control roles assignment sequence obtained with the data-driven MCDM CCRCs schedule.}
    \label{fig: sched_IPC}
    \vspace{-0.7cm}
\end{figure}

According to the \textit{Data-driven re-scheduling}, the sequence of control roles assigned to each IPC is shown in Fig.~\ref{fig: sched_IPC}. In addition to IPC-E, which is constrained to AC-GFM by its operating principle, the other IPCs of the DC-grid 2 also maintain fixed control roles — one in AC-GFM and the other in DC-GFM. The IPCs of the DC-grid 1, however, are the ones subjected to control role switches. The data-driven MCDM was solved by setting the maximum number of allowed simultaneous control switches equal to 1 ($\gamma^* = 1$). This constraint is satisfied in all cases except one, where it is necessary to relax it and allow up to 2 simultaneous switches.  

\begin{figure}[h!]
    \centering
    \includegraphics[width=0.9\linewidth]{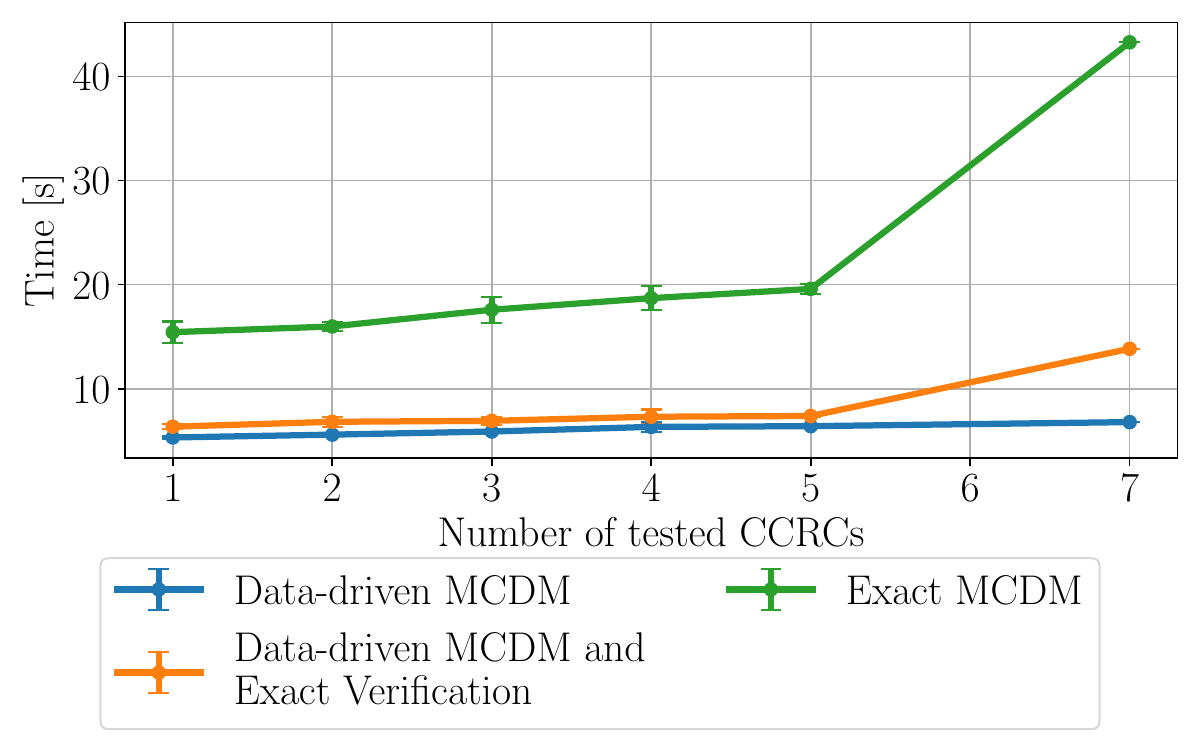}
    \caption{Computing time}
    \label{fig: comptime}
    \vspace{-0.7cm}
\end{figure}

\vspace{-0.3cm}
The computational performance of the proposed data-driven MCDM is evaluated in comparison to the \textit{exact} MCDM. During the execution of the MCDM for the 15-minute operating points over a 24-hour period, the computing time required for solving each instance was recorded for both methods. The PC used is equipped with an Intel Core i7-11390H (4 cores / 8 threads, 3.6 GHz), 16 GB RAM. Fig.~\ref{fig: comptime} illustrates the computing time as a function of the number of CCRC alternatives (i.e., the CCRCs in $\mathcal{R}'_\cap$) considered per instance. For the data-driven MCDM, the total computing time is further decomposed into the time required solely for solving the MCDM using the \textit{surrogate} models, and the additional time needed to verify the solution with the \textit{exact} model. On average, the data-driven MCDM achieves a computing time reduction of 59.8\% compared to the \textit{exact} MCDM. Moreover, unlike the \textit{exact} MCDM, its computing time is considerably reduced for any number of CCRC alternatives at each execution. This characteristic makes it a promising approach for application to larger systems with a higher number of IPCs and CCRCs.
\vspace{-0.5cm}

\vspace{-0.2cm}
\section{Conclusion}
This paper presents a tool for supporting System Operators in enhancing small-signal stability performance of hybrid AC/DC grids, by performing an online IPC control role assignment. The tool is based on a data-driven MCDM algorithm. A comprehensive methodology for training accurate data-driven surrogate models for stability assessment and stability indicators calculation is presented, which does not require sharing the IPC control algorithms but only electrical schemes. A strategy to enhance the scalability of the data-driven MCDM  is introduced, able to minimize the number of  CRCs required to operate the system with good dynamic performance. The proposed models demonstrate high accuracy and the implementation of the data-driven MCDM shows high computational efficiency.  
\vspace{-0.2cm}

\section*{Acknowledgments}
This work was supported by FEDER/Ministerio de Ciencia, Innovación y Universidades - Agencia Estatal de Investigación, Project PDC2022-133226-I00 Total ACDC


\bibliographystyle{IEEEtran}

\newpage

 





\end{document}